\documentclass[aps,prd,10pt,twocolumn,superscriptaddress,showpacs]{revtex4-1}

\pdfoutput=1
\usepackage{hyperref}
\usepackage{graphicx}
\usepackage{amsfonts,amsmath,amssymb,bm,bbm}
\usepackage{color}
\usepackage{enumitem}
\usepackage{soul,xcolor}
\setstcolor{red}

\newcommand{\blue}[1]{\textcolor{blue}{#1}}

\newcommand{\sav}{SA$\nu$}
\newcommand{\eav}{EA$\nu$}

\newcommand{\beq}{\begin{equation}}
\newcommand{\eeq}{\end{equation}}

\hypersetup{
    pdfnewwindow=true,      
    colorlinks=true,       
    linkcolor=blue,          
    citecolor=blue,        
    filecolor=blue,      
    urlcolor=blue        
}

\def\bi{\begin{itemize}[noitemsep,leftmargin=*]}
\def\ei{\end{itemize}}

\begin{document}

\title{Solar Atmospheric Neutrinos: A New Neutrino Floor for Dark Matter Searches}

\author{Kenny C. Y. Ng}
\email{chun-yu.ng@weizmann.ac.il}
\affiliation{
Department of Particle Physics and Astrophysics, Weizmann Institute of Science, Rehovot, Israel}
\affiliation{Center for Cosmology and AstroParticle Physics (CCAPP), Ohio State University, Columbus, OH 43210}
\affiliation{Department of Physics, Ohio State University, Columbus, OH 43210}

\author{John F. Beacom}
\email{beacom.7@osu.edu}
\affiliation{Center for Cosmology and AstroParticle Physics (CCAPP), Ohio State University, Columbus, OH 43210}
\affiliation{Department of Physics, Ohio State University, Columbus, OH 43210}
\affiliation{Department of Astronomy, Ohio State University, Columbus, OH 43210} 

\author{Annika H. G. Peter}
\email{apeter@physics.osu.edu}
\affiliation{Center for Cosmology and AstroParticle Physics (CCAPP), Ohio State University, Columbus, OH 43210}
\affiliation{Department of Physics, Ohio State University, Columbus, OH 43210}
\affiliation{Department of Astronomy, Ohio State University, Columbus, OH 43210} 

\author{Carsten Rott}
\email{rott@skku.edu}
\affiliation{Department of Physics, Sungkyunkwan University, Suwon 440-746, Korea}

\date{13 November 2017}

\begin{abstract}
As is well known, dark matter direct detection experiments will ultimately be limited by a ``neutrino floor," due to the scattering of nuclei by MeV neutrinos from, e.g., nuclear fusion in the Sun.  Here we point out the existence of a new ``neutrino floor" that will similarly limit indirect detection with the Sun, due to high-energy neutrinos from cosmic-ray interactions with the solar atmosphere.  We have two key findings.  First, solar atmospheric neutrinos $\lesssim 1$ TeV cause a sensitivity floor for standard WIMP scenarios, for which higher-energy neutrinos are absorbed in the Sun.  This floor will be reached once the present sensitivity is improved by just one order of magnitude.  Second, for neutrinos $\gtrsim 1$ TeV, which can be isolated by muon energy loss rate, solar atmospheric neutrinos should soon be detectable in IceCube.  Discovery will help probe the complicated effects of solar magnetic fields on cosmic rays.  These events will be backgrounds to WIMP scenarios with long-lived mediators, for which higher-energy neutrinos can escape from the Sun.
\end{abstract}



\maketitle


\section{Introduction}
\label{sec:introduction}

Numerous astrophysical and cosmological observations show that most of the matter in the universe has no apparent electromagnetic interactions, and hence is called dark matter~(DM)~\cite{Jungman:1995df, Bertone:2004pz, Strigari:2013iaa}.  Identifying the particle nature of DM is important for understanding what lies beyond the standard models of cosmology and of particle physics.

Weakly Interacting Massive Particles~(WIMPs)~\cite{Steigman:1984ac}, which can be produced with the correct abundance as a thermal relic of the early universe~\cite{ZELDOVICH1965241, Steigman:2012nb}, are a popular DM candidate.  WIMPs can be probed through annihilation signals seen by high-energy astrophysical observatories~(indirect detection)~\cite{Conrad:2014tla, Gaskins:2016cha}, production at colliders detected by missing energy~\cite{Goodman:2010ku, Abdallah:2015ter}, and by the elastic scattering of nuclei in underground experiments~(direct detection)~\cite{Baudis:2012ig, Peter:2013aha, Gluscevic:2014vga}.  

As direct detection experiments improve in sensitivity, they will reach a ``neutrino floor," due to nuclear recoils induced by neutrinos from the Sun, cosmic supernovae, and Earth's atmosphere~\cite{Cabrera:1984rr, Monroe:2007xp, Strigari:2009bq, Gutlein:2010tq, Harnik:2012ni, Billard:2013qya, Ruppin:2014bra}.  This is an irreducible background that cannot be shielded like other typical backgrounds in underground experiments. Once these neutrinos are detected, the sensitivity to DM scattering can improve at best with the square root of exposure.  Importantly, current detectors are almost reaching this floor in some parameter space~\cite{Akerib:2016vxi, Aprile:2016swn, Akerib:2015rjg, Tan:2016zwf}.  

The scattering of DM with nuclei can also be probed by DM capture in the Sun~\cite{1985ApJ296679P, Krauss:1985ks, Silk:1985ax, Peter:2009mk}.  As the Sun moves in the Galactic DM halo, DM particles can scatter with the Sun, lose some of its kinetic energy such that the final velocity is below the escape velocity~(gravitational capture).  The DM particles continue to lose energy through scattering and eventually accumulate at the core of the Sun.  The accumulated DM can annihilate into various channels and produce neutrinos that can be searched by neutrino telescopes.  
While for spin-independent~(SI) scattering, the best sensitivity comes from direct detection experiments, for spin-dependent~(SD) scattering, the sensitivity of solar DM searches with neutrino telescopes can be better than that of direct detection experiments, depending on the annihilation channel~\cite{Choi:2015ara, Aartsen:2016zhm}.  
At present, the dominant background for solar DM searches is Earth atmospheric neutrinos~(\eav), and thus the sensitivity improves with the square root of exposure.   

However, as solar DM searches become more sensitive, they too will face a ``neutrino floor,'' caused by solar atmospheric neutrinos~(\sav~\cite{Moskalenko:1991hm, Seckel:1991ffa, Moskalenko:1993ke, Ingelman:1996mj}), produced by cosmic-ray interactions with the Sun.  The \sav\ background is especially troublesome in the sense that it cannot be distinguished from the signal using the arrival direction.  This, together with the considerable flux uncertainty and poor energy resolution of muon neutrinos~(the most important search channel), make the \sav\ background far more difficult to eliminate than the smoothly distributed and well-measured \eav\ background.  Also due to the flux uncertainty, the DM sensitivity cannot improve once \sav\ is detectable.  Thus, the SA${\nu}$ constitutes a \emph{hard} floor.  If the SA${\nu}$ flux can be predicted well, or distinguished by the spectral shape, then the sensitivity can improve with the square root of exposure.  In this case, the sensitivity floor is \emph{soft}.  

In this paper, we first consider SA${\nu}$ as an interesting signal and discuss its detectability. 
We then consider SA${\nu}$ as a background to the solar DM search, and calculate the neutrino floor.  (Some of our preliminary results were presented in conferences, e.g., Refs.~\footnote{https://meetings.wipac.wisc.edu/IPA2015/program, https://fermi.gsfc.nasa.gov/science/mtgs/symposia/2015/}).  In Sec.~\ref{sec:sav}, we review the \sav\ flux and calculate the detection prospects.  In Sec.~\ref{sec:neutrino_floor}, we review the neutrino flux from DM captured in the Sun and calculate the DM sensitivity floor caused by \sav.  We also discuss its implications for non-minimal WIMP models.  We aim for a precision of a factor of $\sim 2$, considering the uncertainties involved.  We conclude in Sec.~\ref{sec:conclusion}, where we also discuss how these uncertainties can be reduced.


\section{Solar Atmospheric Neutrino Fluxes and Detection}
\label{sec:sav}


\begin{figure}[t]
\includegraphics[width=8.5cm]{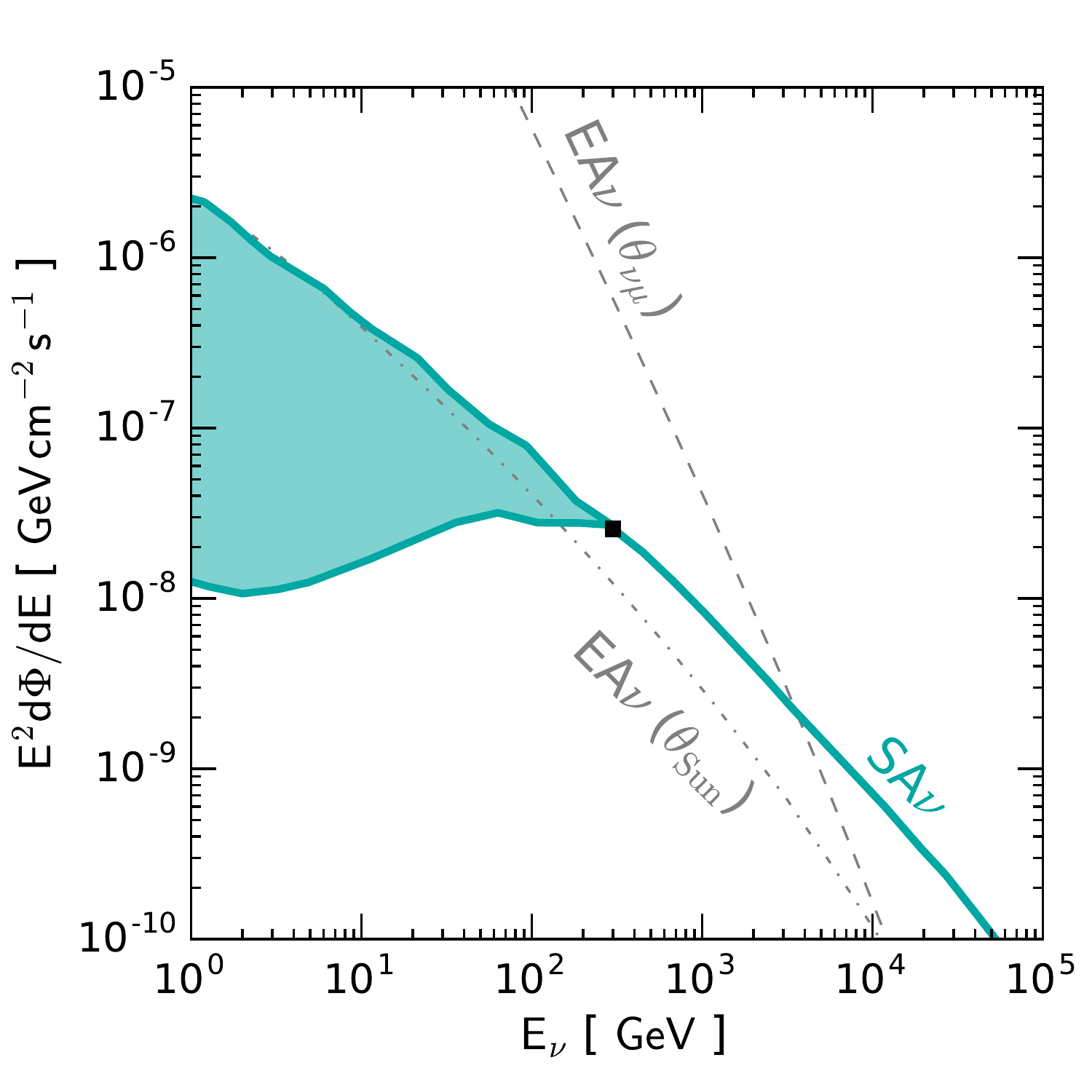}
\caption{The \sav\ flux spectrum.  Below 300 GeV, we use the SSG1991 models~\cite{Seckel:1991ffa} (Upper: {\it Naive}; Lower: {\it Nominal}); above 300 GeV, we use the IT1996 model~\cite{Ingelman:1996mj}.  All are shown within the angular cone of the Sun ($\theta_{\rm Sun}$).  We also show the \eav\ flux spectrum within $\theta_{\rm Sun}$ and within the neutrino-muon separation angle~($\theta_{\nu\mu}$).
}
\label{fig:neutrino_flux}
\end{figure}

\begin{figure}[t]
\includegraphics[width=8.5cm]{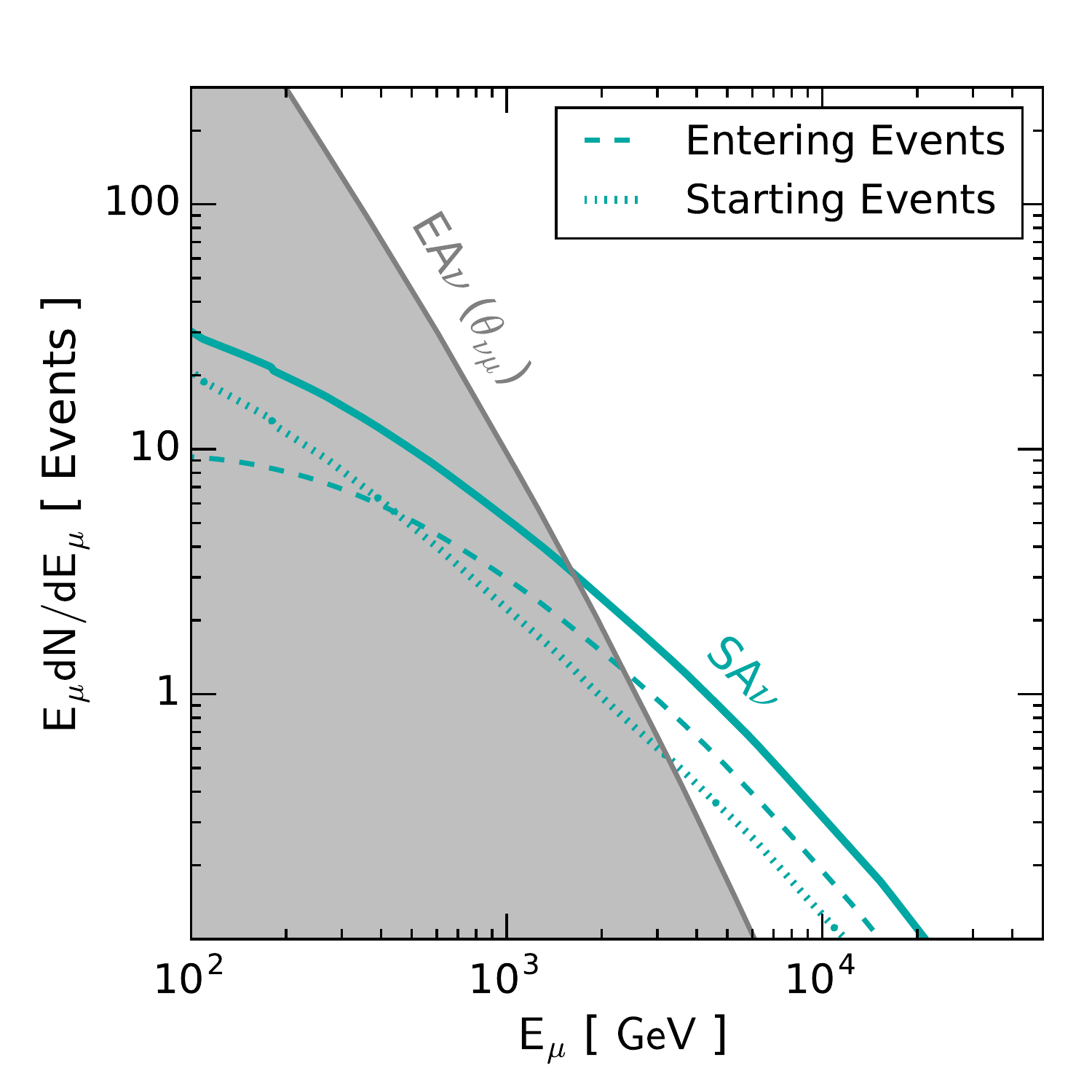}
\caption{ The total muon spectrum~(starting + entering) of the \sav\ flux~(nominal model) for IceCube with 10 years of live time.  
Also shown is the muon spectrum from \eav\ background within the neutrino-muon opening angle.  The \sav\ signal exceeds the background at $E_{\mu} \gtrsim2\,{\rm TeV}$.  }
\label{fig:sav_muspec}
\end{figure}

\subsection{Solar Atmospheric Neutrino Flux}

In this subsection, we review the \sav\ flux calculation and define the model we use.  We focus on muon neutrinos and their charged-current interactions, as the directionality provided by the final-state muons is crucial for detection and for background reduction, especially in the TeV range.

Cosmic rays entering the solar atmosphere undergo hadronic interactions and produce secondary particles, such as charged pions and kaons.  These secondary particles can then decay into neutrinos, thus lead to the production of \sav~(also from $p\gamma$ interactions at $\gtrsim1$\,PeV~\cite{Andersen:2011dz}).  The process is similar to the production of \eav~\cite{Gaisser:2002jj}.  For both cases, the thick-target limit is appropriate for the cosmic rays, which means that the column density is high enough that interactions are very likely.  Therefore, the \sav\ and \eav\ are naively expected to have comparable intensity~(flux per solid angle).  However, there are some important differences. 

If solar magnetic fields are ignored, then the most important difference is at high neutrino energies.  In Earth's atmosphere, pions~(kaons) above about 100\,GeV~(800\,GeV) undergo significant hadronic scattering before decay~\cite{Gaisser:2002jj}, lowering the energy of their neutrinos and thus steepening their spectrum compared to the cosmic-ray spectrum.  The Sun's atmosphere is thinner, so this steepening does not occur until one to two orders of magnitude higher in energy~\cite{Moskalenko:1991hm, Seckel:1991ffa, Moskalenko:1993ke, Ingelman:1996mj}.  This difference makes the \sav\ spectrum both higher and harder in the TeV range, which is an important distinction between \sav\ and \eav.  

Magnetic fields are important at lower energies.  Cosmic-ray propagation in the solar system is affected by solar magnetic fields carried by the solar wind; magnetic fields near the solar surface can also affect the propagation of cosmic rays and their charged secondaries. These effects were modeled by Seckel, Stanev, and Gaisser~(SSG1991~\cite{Seckel:1991ffa}).  In their {\it Nominal} model, the rate of cosmic rays interacting with the solar atmosphere is reduced due to reflection by the magnetic flux tubes in the solar surface.  This leads to a strong suppression of the neutrino flux at low energies.  In their {\it Naive} model, where magnetic effects are ignored, the \sav\ intensity is indeed comparable to the \eav\ intensity near $\sim 1$\,GeV.  At sufficiently high energies, magnetic effects should diminish.  In the SSG1991 models, this transition occurs at about 300 GeV neutrino energy, though the value is theoretically quite uncertain.  At lower energies, the spread between the SSG1991 models gives some indication of the uncertainty.  The corresponding gamma-ray fluxes lie between these two extremes~\cite{Orlando:2008uk, Abdo:2011xn, Ng:2015gya}.  We use the SSG1991 models up to 300\,GeV.  

At higher energies, the uncertainties are expected to be less, but \blue{could} be non-negligible.  For neutrino energies above 300\,GeV, we use the model from Ingelman and Thunman~(IT1996~\cite{Ingelman:1996mj}).  The IT1996 model assumes zero magnetic fields, and is consistent with the {\it Naive} model of SSG1991 above $\sim 100$\,GeV.  We caution that it is not clear how much magnetic fields can affect the neutrino production at $\sim 1$\,TeV, the most relevant energy range for \sav\ detection, and we comment further in Sec.~\ref{sec:sav_signal}. 

We take into account neutrino mixing.  As shown in Refs.~\cite{Fogli:2006jk}, there are both vacuum-mixing and matter effects.  However, these effects are largely washed out after combining neutrino and anti-neutrinos, integrating over the production region, and using wide energy bins.  The final muon neutrino flux is thus roughly a factor of $\simeq 0.5$ less than that at production, similar to vacuum mixing alone, where $1:2:0$ transforms to nearly $1:1:1$.  For simplicity, given the other large uncertainties, we simply reduce the total \sav\ muon neutrino flux by this factor.

For the \eav\ model, we use the all-sky averaged intensity from Ref.~\cite{Honda:2015fha}, and the parametric form in Ref.~\cite{Sinegovskaya:2014pia} to extrapolate to high energies, after matching the normalization.  We ignore neutrino mixing for the \eav, which would reduce the flux by a factor of 2 at low energies and would be negligible at high energies~\cite{Aartsen:2014yll}, where we are most interested. The \eav\ intensity also changes with zenith angle~\cite{Honda:2006qj}, but is only a $\sim 50$\% effect for the most important energies and directions considered here.  We neglect this variation, in keeping with our precision goal of a factor of $\sim 2$.  

Figure~\ref{fig:neutrino_flux} shows the predicted \sav\ flux after mixing, integrated over the angular size of the Sun.  We have joined the SSG1991 and IT1996 fluxes at 300 GeV.  We also show the corresponding \eav\ flux within the angular size of the Sun, with half angle~$\theta_{\rm Sun} = 0.27^\circ$. As described above, in the same solid angle, the \eav\ flux becomes smaller and steeper than the \sav\ flux at high energies.  

However, the actual relevant \eav\ background should be given by the flux within the neutrino-muon separation angle, $\theta_{\nu\mu}\simeq 1^{\circ}\sqrt{1\,{\rm TeV}/E_{\nu}}$~\cite{Gaisser:1990vg, Aartsen:2013vja}.  This is the mean angle between the incoming neutrinos and the outgoing muons, after the neutrino-quark charged-current interactions.  It is therefore an intrinsic limitation to the best possible neutrino angular resolution if only the final state muons are observed, and is independent of the detector technology.  As shown in Fig.~\ref{fig:neutrino_flux}, even in this case, the \sav\ flux exceeds the \eav\ background above a few TeV.  


\subsection{Neutrino Detection}
\label{sec:neutrino_detection}

In this subsection, we discuss the detection of muon neutrinos from the Sun with neutrino telescopes.  We adopt the ``theorist's'' or ideal approach to estimate the best possible scenario.  In a realistic case, background reduction and threshold effects reduce the signal efficiency, which are encoded in the effective areas provided by experimental collaborations.  These effective areas are thus analysis-dependent, and could be improved.  The ideal approach is necessary because we want to separate events by muon energy, which is not possible in the effective-area approach.  We comment on the differences between the ideal and the realistic cases below.

As noted, we focus on muon neutrinos and the tracks they produce in charged-current interactions.  We combine neutrinos and antineutrinos.  The muon energy at birth, $E_{\mu}$, is related to the neutrino energy, $E_{\nu}$, by $ E_{\mu} = E_{\nu} (1 - y )$, where $ y $ is the inelasticity parameter~\cite{Gandhi:1998ri, Gandhi:1995tf}. For simplicity, we assume a fixed value of $y = 0.4$ throughout our energy range of interest.  We neglect neutrino absorption in Earth, which becomes important only above $\sim40$\,TeV for neutrinos that cross the diameter (and $\sim1$\,PeV for neutrinos that travel from the Sun to IceCube~\cite{Gandhi:1995tf}).

Muons can be produced inside the detector~(starting events), or outside and then enter the detector after propagation~(entering events).  For starting events, the muon spectrum is
\begin{equation}
\frac{dN^{\rm sta}}{dE_{\mu}} \simeq {N_{A} \rho V T}\frac{1}{1-y} \left[\frac{d\Phi}{dE_{\nu}}(E_{\nu}) \sigma(E_{\nu}) \right]_{E_{\nu} = \frac{E_{\mu}}{(1-y)}}\,,
\end{equation}
where $d\Phi/dE_{\nu}$ is the neutrino flux , $\sigma$ is the interaction cross section~\cite{Gandhi:1998ri, Gandhi:1995tf},  $N_{A} = 6.02\times 10^{23}\,g^{-1}$ is the Avogadro number, $\rho \simeq 1\,{\rm g\,cm^{-3}}$ is the density, $V$ is the fiducial volume of the detector, and $T$ is the effective exposure.  The muon energy is taken to be its birth energy.  To reduce backgrounds from atmospheric muons, we consider only upgoing events. 
The effective exposure for the Sun is thus taken to be half the detector live time.

For entering muons, taking into account energy loss, the spectrum is ~\cite{Gaisser:1990vg, Kistler:2006hp}
\begin{equation}
\frac{dN^{\rm ent}}{dE_{\mu}} \simeq \frac{N_{A} \rho A T}{ \rho \left(\alpha  + \beta E_{\mu}\right) } \int_{ \frac{E_{\mu}}{1-y} }^{\infty}  dE_{\nu} \frac{d\Phi}{dE_{\nu}}(E_{\nu}) \sigma(E_{\nu})\, ,
\end{equation}
where $A$ is the geometric detector area, $\alpha = 2.0 \times 10^{-6}\,{\rm TeV\, cm^{2}\, g^{-1}}$, and $\beta = 4.2 \times 10^{-6}\, {\rm cm^{2}\,g^{-1} }$~\cite{Lipari:1991ut, Dutta:2000hh}.  The muon energy is that when the muon enters the detector.

We consider two idealized experimental setups that roughly correspond to Super-Kamiokande~(Super-K) and IceCube.  They cover the range of a small, low-threshold detector and a large, high-threshold detector, and are representative of similarly sized future detectors.  For Super-K, we use $V \simeq 2\times 10^{4}\,{\rm m^{3}}$ and approximate the geometric area to be $A \simeq 780\,{\rm m^{2}}$.  For IceCube, we use $V \simeq 10^{9}\,{\rm m^{3}}$ and $A \simeq 10^{6}\,{\rm m^{2}}$.  We discuss the effect of a more realistic setup below, and future detectors in Sec.~\ref{sec:conclusion}. 

\subsection{Solar Atmospheric Neutrinos as Signal}
\label{sec:sav_signal}

In this subsection, we consider the \sav\ as a  signal.  For this case, only an IceCube-sized detector is relevant.  We then discuss the implications of detecting the Sun as a high-energy astrophysical neutrino source. 

Figure~\ref{fig:sav_muspec} shows the muon spectrum of the \sav\ signal compared to the \eav\ background, following the procedure described above, with 10 years of IceCube live time.   At this energy range, the difference between naive and nominal model is small; we use the naive model in this part for concreteness.  
At low energies, $\sim$\,100GeV, the \eav\ background is dominant.  The background decreases rapidly, largely due to the decreasing neutrino-muon angle, and eventually falls below the \sav\ flux.  Therefore, detection of the \sav\ signal critically depends on isolating high-energy events.  Fortunately, this can be done with muons with energy $> 1\,$TeV, which is above the minimum-ionizing regime.  In this regime, the muon energy loss become radiative~\cite{Aartsen:2013vja}, which can be used to distinguish muons above 1\,TeV, as demonstrated in Ref.~\cite{Aartsen:2016exj}.  We find the integrated number of events above 1\,TeV to be 4.5 and 4.1 for \sav\ and \eav, respectively.  Above a slightly higher energy, the signal would decrease, but the background would decrease more.  This suggests that IceCube and KM3NeT~\cite{Adrian-Martinez:2016fdl} are sensitive to the \sav\ signal. 

If TeV muon events are detected from the Sun, we note that they can be distinguished from those from solar DM in standard WIMP scenarios.  As described below, neutrinos above about 100 GeV produced in the solar core are absorbed as they leave the Sun.  Therefore, if an excess of $>1$\,TeV muons is seen from the Sun, they are likely to be \sav\ events.  As a result, we do not count these events when calculating the neutrino sensitivity floor for standard WIMPs.  We further comment on non-minimal DM scenarios below. 

Given that \sav\ could potentially be detected as a signal, it is important to discuss the uncertainty of the \sav\ flux and the implications of a detection.  Most of the inputs of the \sav\ flux calculation, such as the primary cosmic-ray flux, solar matter distribution, and neutrino mixing parameters, are well constrained.  The most uncertain aspect of the \sav\ calculation is the effect of solar magnetic fields.  Theoretically, the inclusion of their effects is challenging due to the complicated solar coronal and photospheric magnetic fields.   From cosmic-ray shadow measurements, there is evidence that coronal fields~\cite{Amenomori:2013own} can affect the propagation of $\sim 10\,{\rm TeV}$ cosmic rays, which is the most relevant energy range for IceCube.  Typically, solar magnetic fields are expected to reduce the rate of cosmic-ray interactions by magnetic reflection of incoming cosmic rays~\cite{Seckel:1991ffa}; thus the neutrino production rate is reduced.  This picture may be more complicated at the IceCube energy range, where neutrino absorption in the Sun is important.   A detection or a constraint on the \sav\ flux will be important to understand the effect of magnetic fields.

One can also study cosmic-ray interactions with the Sun through gamma-ray observations.  Gamma rays are readily absorbed by the Sun.  The no-magnetic-field scenario therefore corresponds to minimal gamma-ray production~\cite{Zhou:2016ljf}.  Observations with Fermi~\cite{Orlando:2008uk, Abdo:2011xn, Ng:2015gya} show that the gamma-ray flux is much larger than the no-magnetic-field case, possibly up to 100\,GeV.  This suggests that magnetic fields can boost gamma-ray production, and affect cosmic-ray primaries up to at least 1\,TeV.  
Above the energy range that magnetic fields can be ignored, gamma-ray production is expected to be suppressed.   As a result, even  limits on the flux of TeV gamma rays from the Sun by HAWC~\cite{Abeysekara:2013tka} and LHAASO~\cite{He:2016del} would be an important clue.  This, together with the detection of \sav\ by IceCube, would be important for normalizing cosmic-ray interaction rates with the Sun and disentangling magnetic-field effects.  

Lastly, it is  important to emphasize that the detection of the Sun as a high-energy neutrino point source would by itself an important milestone for neutrino astronomy, especially given that sources have yet to be identified for the IceCube events.  The Sun is also conveniently observable for neutrino telescopes at both hemispheres, so it could in principle be a flux calibration source for IceCube, KM3NeT, and their successors.

\begin{figure*}
\includegraphics[width=8.5cm]{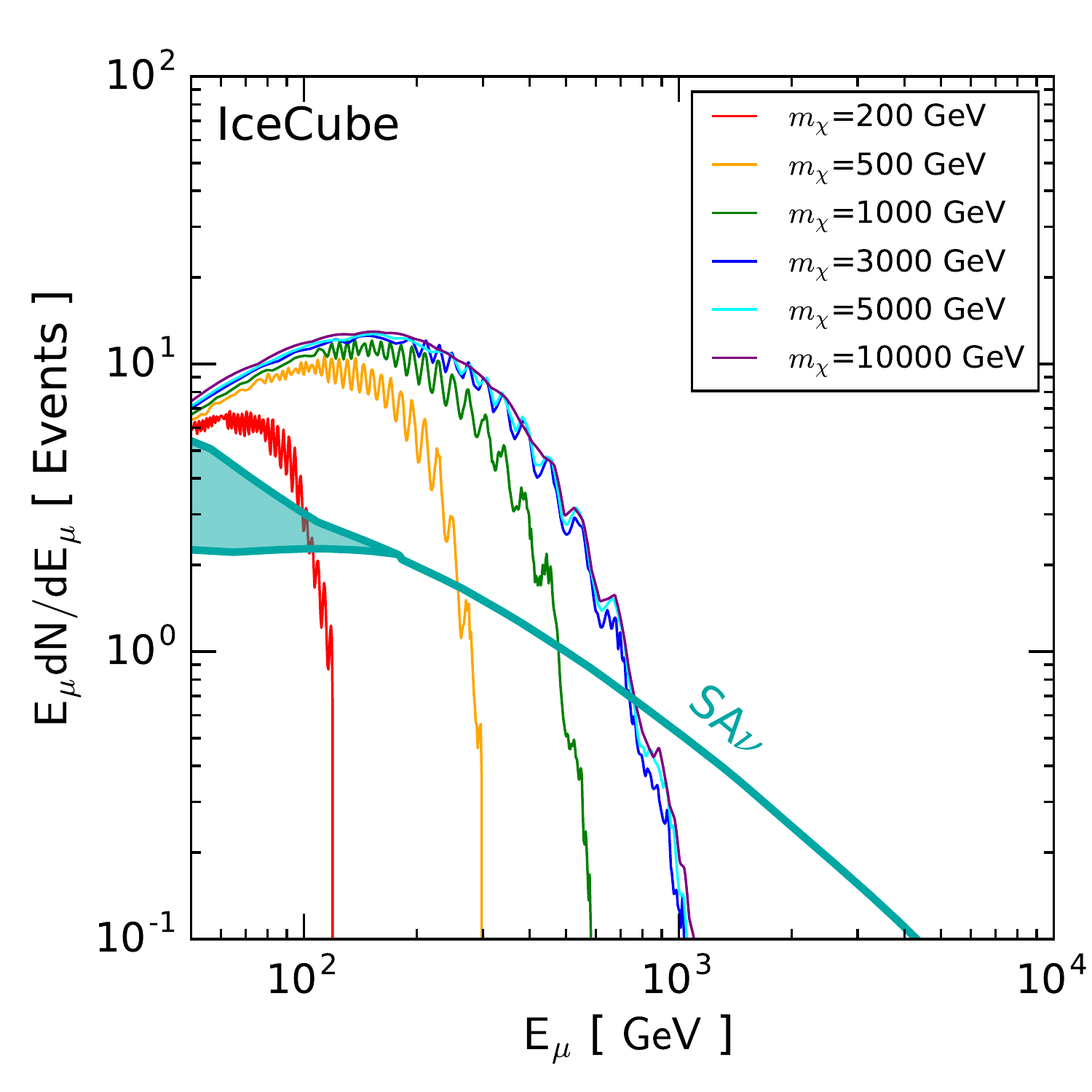}
\includegraphics[width=8.5cm]{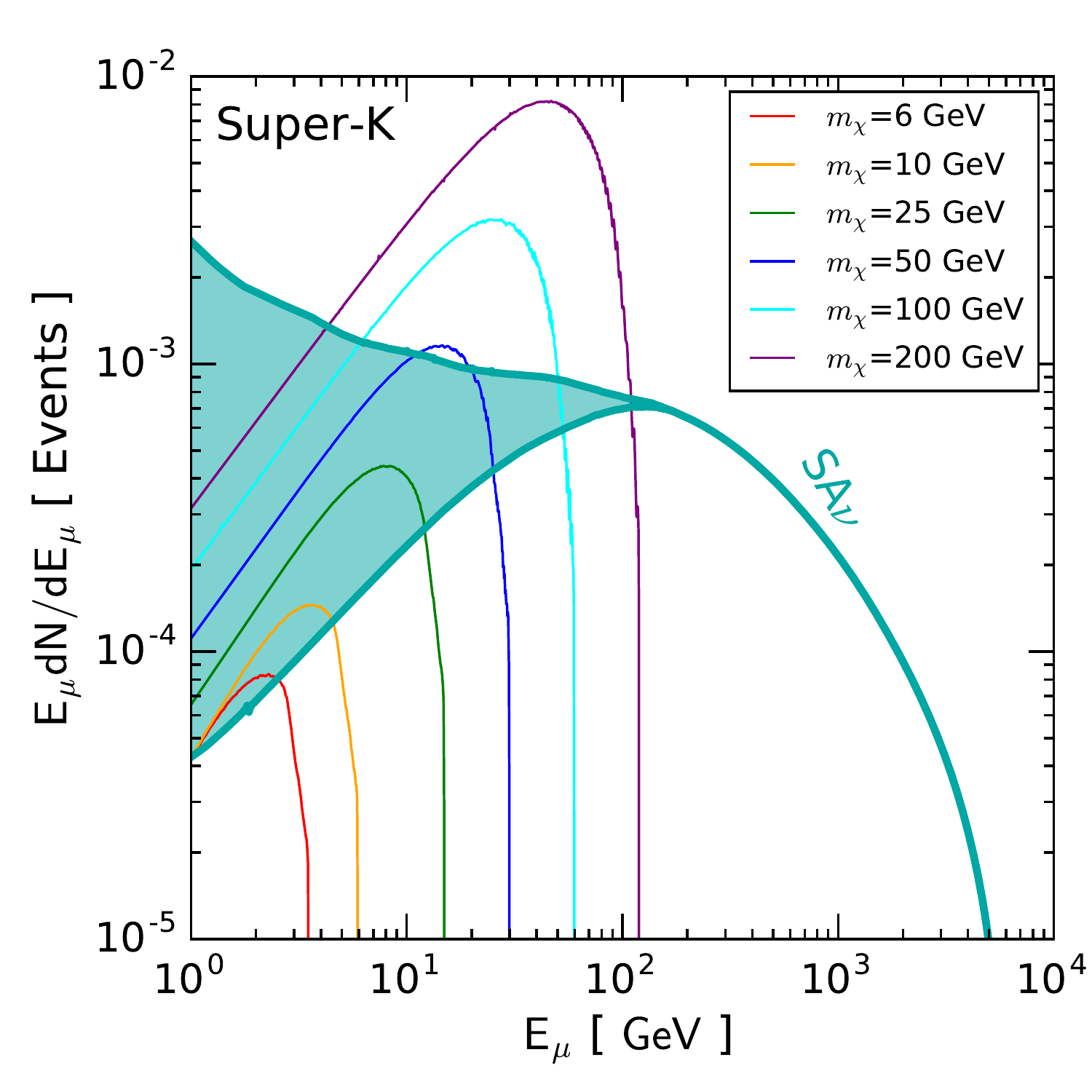}
\caption{
{\bf Left:} The total muon spectrum of the \sav\ in IceCube with 1 year of live time, compared with several DM spectra in the $\tau{\bar \tau}$ channel, obtained with {\sc WimpSim}~\cite{Blennow:2007tw}.  For high DM masses, the spectra become indistinguishable due to neutrino absorption in the Sun.  For presentation, the DM annihilation rate for different masses is taken to have a fixed value of $10^{19}\,{\rm s^{-1}}$~(see text for details). 
{\bf Right:} The same, but for Super-K, and the annihilation rate is $10^{20}\,{\rm s^{-1}}$.   }
\label{fig:dm_spec}
\end{figure*}

\section{Solar DM sensitivity floor}
\label{sec:neutrino_floor}

\subsection{Neutrinos from Solar WIMP DM}

In this subsection, we review the calculation of neutrino flux from WIMP DM annihilation in the Sun. 
The process of DM capture and annihilation in the Sun is well studied~\cite{Gould:1991hx, Peter:2009mk, Rott:2011fh, Danninger:2014xza}.  The time evolution of the DM number density~$N_{\chi}$ in the Sun is
\beq
  \frac{d}{dt}N_\chi=\Gamma_{\rm cap}-C_{\rm ann}N_\chi^2\, ,
\eeq
where $\Gamma_{\rm cap}$ is the capture rate of DM in the Sun and $C_{\rm ann}$ is the annihilation coefficient.  We ignore the evaporation term, which is only relevant below $\sim4$\,GeV~\cite{Busoni:2013kaa, Kouvaris:2015nsa}. 
For typical parameters, equilibrium is achieved~\cite{Peter:2009mk}.  Hence, the annihilation rate, $\Gamma_{\rm ann}$, is related to the capture rate, $\Gamma_{\rm ann} = C_{\rm ann}N_\chi^2/2 \simeq \Gamma_{\rm cap}/2$.

The capture rate, $\Gamma_{\rm cap}$, depends on the DM-nucleon cross section and the DM mass $m_{\chi}$, and is proportional to the probability that a DM particle's velocity falling below the Sun's escape velocity after the scattering~(gravitational capture).  We calculate the capture rates, and hence the annihilation rates, using {\sc DarkSUSY}~\cite{Gondolo:2004sc} version 5.1.3 with default settings, which performs a numerical integration using the prescription given in Ref.~\cite{Gould:1987ir}.  Unlike other indirect detection methods, such as DM annihilation in the Galactic Center, the capture rate in the Sun is not very sensitive to astrophysical uncertainties~\cite{Danninger:2014xza,Choi:2013eda}.    

The differential neutrino flux, $d\Phi_{\nu}/dE$ is
\beq
\frac{d\Phi}{dE_{\nu}}=\frac{\Gamma_{\rm ann}}{4\pi D_\oplus^2} \frac{d\widetilde{N}}{dE_{\nu}},
 \label{eq:flux}
\eeq
where $d\widetilde{N}/dE$ is the neutrino spectrum per annihilation (with all mixing effects included) and $D_\oplus \simeq 1.5\times 10^{8}\,{\rm km}$ is the distance to the Sun.

We obtain the neutrino spectrum per annihilation using {\sc WimpSim}~\cite{Blennow:2007tw} version 3.03~(available in \footnote{http://wimpsim.astroparticle.se/results.html}), which takes into account both neutrino absorption in the Sun and flavor evolution from production to the Earth~\cite{Crotty:2002mv, Cirelli:2005gh, Lehnert:2007fv, Baratella:2013fya}. (The latter can be seen from the ``wiggles'' in the spectra.)
We also ignore the very-low-energy neutrinos  from DM annihilation in the Sun~\cite{Rott:2012qb, Bernal:2012qh}.
The neutrino spectra depends on the underlying DM models.  To discuss our results in a model-independent manner, we consider two cases, where DM annihilates into $\tau{\bar \tau}$ and $b {\bar b}$ with 100\% branching fraction, respectively.  Both $\tau{\bar \tau}$ and $b {\bar b}$ are unstable; they decay, or hadronize and then decay into various final products including neutrinos. 
These two channels are typically used to represent hard and soft spectral shapes.    

Figure~\ref{fig:dm_spec} shows the total DM muon spectra for the $\chi\chi\rightarrow\tau{\bar \tau}$ channel as well as the \sav\ muon spectrum for IceCube and Super-K, respectively. 
For illustration, the input neutrino spectra are chosen to have the same annihilation rates, and hence comparable number fluxes. Higher DM masses simply means higher neutrino energies, which is more favorable for detection due to higher neutrino cross section and increased muon range.  This can be clearly seen for Super-K, where the muon event rates increase significantly with DM mass.  Therefore, lower-mass DM requires a larger annihilation rate to yield comparable events rate as high-mass DM.   However, for a given cross section, the capture rate~(thus, annihilation rate) decreases with the DM mass due to a combination of factors, including the decreasing DM number density and the capture kinematics~(see Figure~1 in Ref.~\cite{Rott:2012qb}), the final sensitivity to the scattering cross section turns out to be a weak function of the DM mass for Super-K~(see below).  For IceCube, the sensitivity gain with high-energy neutrinos is hampered by the neutrinos absorption in the Sun during their escape from the core of the star.  This introduces a absorption factor, $\sim e^{-X(E_{\nu})}$, to the neutrino spectrum, where $X(E_{\nu})$ is the optical depth.   The optical depth increases with energy following the neutrino cross section, and approaches unity around a few hundred GeV.  This explains why the muon spectra have similar shapes in IceCube for high DM masses, as they are suppressed by the same factor.  Due to the absorption, the cross section sensitivity also weakens above $\sim$\,TeV~(see below).

\subsection{Indirect Detection Neutrino Floor}

In this subsection, we consider \sav\ as a background to solar DM searches, and we calculate the corresponding sensitivity floor.  

To estimate the neutrino sensitivity floor,  we compare the number of \sav\ background events to the DM signal events by integrating the total~(starting + entering) muon spectrum, 
\beq
N = \int_{E_{\rm min}}^{E_{\rm max}}  \left( \frac{dN^{\rm sta}}{dE_{\mu}} + \frac{dN^{\rm ent}}{dE_{\mu}} \right) \, .
\eeq
The energy range, $E_{\rm min}$ to $E_{\rm max}$, depends on the detector.  For IceCube, we choose $E_{\rm min}$ and $E_{\rm max}$ to be 50\,GeV and 1\,TeV.  The lower bound is chosen to roughly match the main IceCube selection in Ref.~\cite{Aartsen:2016zhm}.  We assume events above 1\,TeV can be identified and isolated by energy loss.  They are not included here as standard WIMP DM cannot produce such neutrinos.  (Including the high-energy neutrinos would cause only a modest difference in our results for the floor, because the \sav\ spectrum is falling.)  For Super-K, $E_{\rm min}$ and $E_{\rm max}$ are chosen to be 1\,GeV and 1\,TeV.  
The precise choice of $E_{\rm max}$ does not change our result by much, due to the small number of events for both \sav\ and DM components.  The choice of $E_{\rm min}$ does affect DM masses that are near the threshold.  Here, we assume neutrino telescopes have no energy information for muon tracks, and so only one energy bin is considered.

For energies below about 200\,GeV, the uncertainty of \sav\ flux is estimated using the {\it Naive} and {\it Nominal} models from SSG1991.  As mentioned above, the uncertainty at higher energies is not clear, given the complicated magnetic field effects on cosmic-ray interactions in the Sun. 
As can be seen from Fig.~\ref{fig:dm_spec}, the uncertainty in the \sav\ flux affects Super-K much more than IceCube.  Integrating the {\it Naive} and {\it Nominal} \sav\ spectra, we obtain about $5-6$ events/yr for IceCube and $0.003-0.007$ events/yr for Super-K.  
Qualitatively, we can see that \sav\ events are not likely to be detectable in Super-K.  This already shows that the neutrino sensitivity floor will not be reached by kiloton scale low-threshold detectors. 
For IceCube, however, even in the case when \eav\ backgrounds can be completely removed, the \sav\ events will ultimately limit the DM search.  

\begin{figure*}
\includegraphics[width=8.5cm]{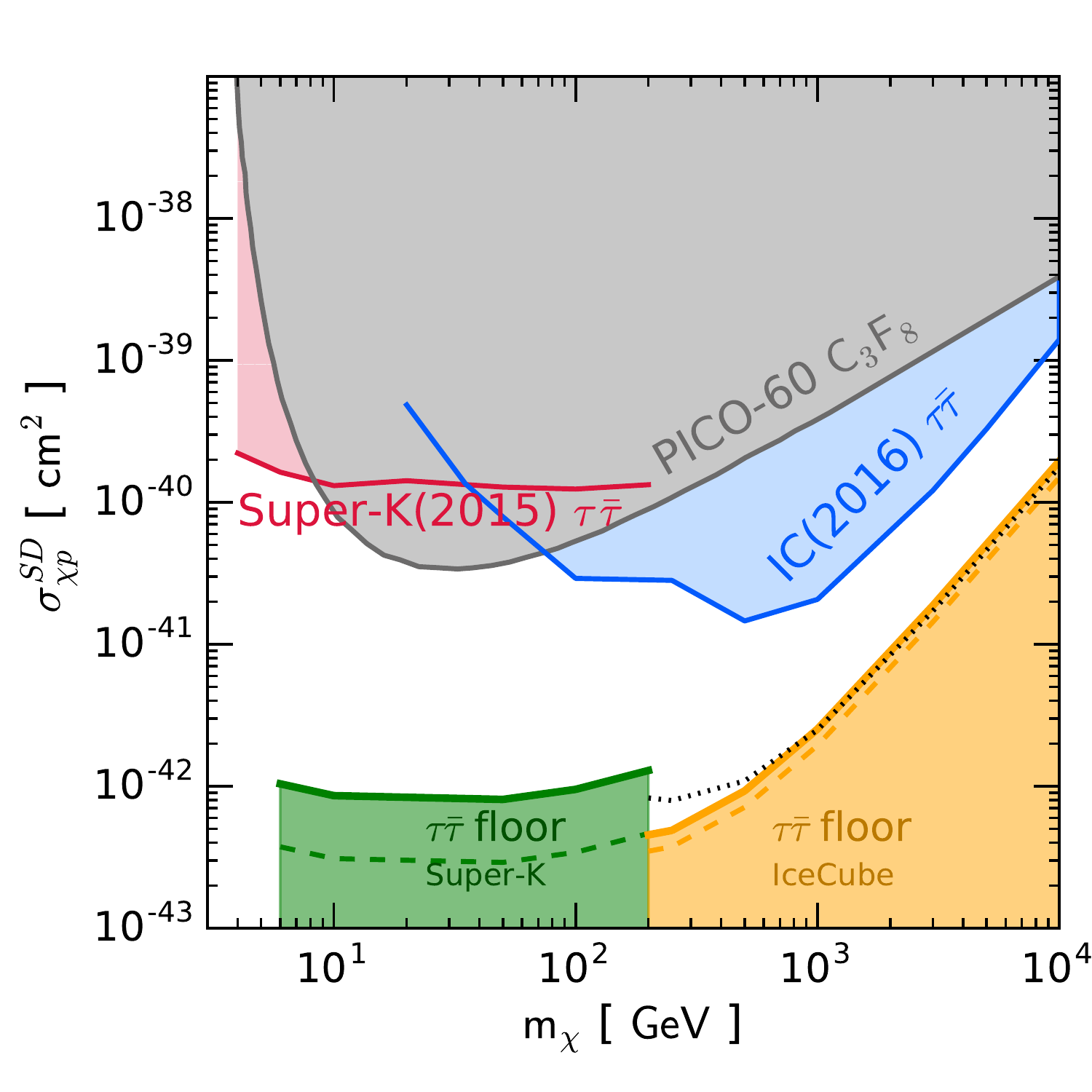}
\includegraphics[width=8.5cm]{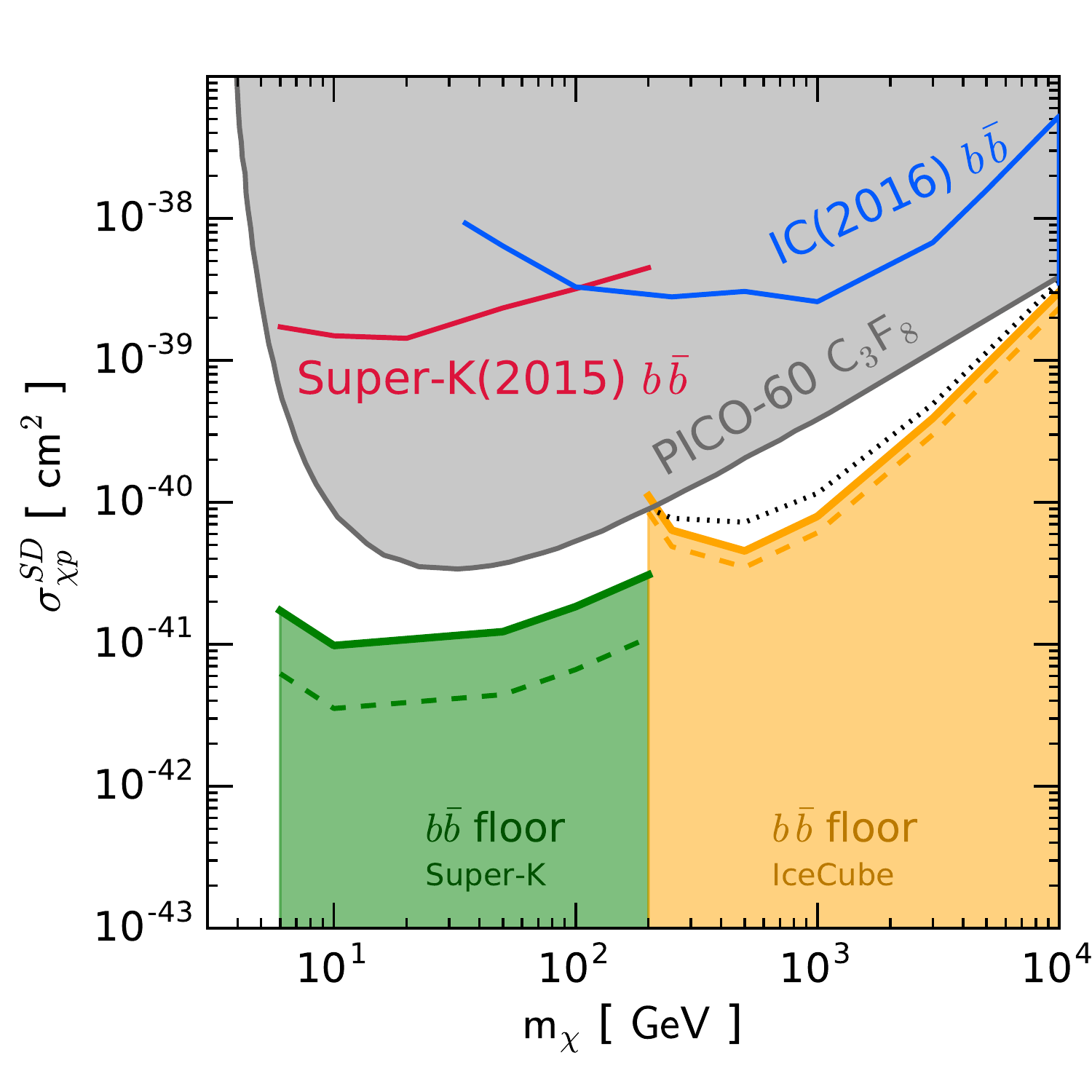}
\caption{
{\bf Left:} 
The indirect detection neutrino sensitivity floors for SD DM cross section for Super-K and IceCube are shown in the bottom.   
We show the case for $\tau{\bar \tau}$ channel, and the solid~(dashed) line corresponds to the {\it Naive}~({\it Nominal}) case.  For comparison, we show, in black dotted line, the floor obtained using the IceCube effective area~\cite{Aartsen:2016zhm} with the {\it Naive} model. 
We also show current indirect detection limits from Super-K~\cite{Choi:2015ara} and IceCube~\cite{Aartsen:2016zhm}; the whitespace between them and the floors shows the remaining parameter space that can be probed by solar DM searches.  For perspective, we also show the direct detection limits from PICO~\cite{Amole:2017dex}.
{\bf Right:} Same as the left, but for $b{\bar b}$ channel.  }
\label{fig:neutrino_floor}
\end{figure*}

To quantitatively calculate the neutrino floor, we find the DM flux that gives an equal number of events to the \sav\ background.  For each DM mass and annihilation channel, this then defines a DM-nucleon cross section.  This is equivalent to assuming the \sav\ events are totally indistinguishable from the DM annihilation events, or there is 100\% uncertainty in the \sav\ flux. 
In principle, the expected number of of \sav\ events can be estimated with an accurate \sav\ model, or inferred from  \sav\ observations at energies $>$\,TeV.  However, given that there is likely appreciable uncertainty in the \sav\ flux, as suggested by gamma-ray data at $\simeq 100$\,GeV, the 100\% uncertainty is reasonable and perhaps optimistic.  
Lastly, we also neglect the \eav, as well as other backgrounds.  Including these would increase the floor. 

Figure~\ref{fig:neutrino_floor} shows the neutrino sensitivity floor for the $\tau{\bar \tau}$ and $b{\bar b}$ channels, which represents hard and soft DM spectra, respectively. We only consider SD cross sections, as direct detection experiments are more efficient at probing the SI cross sections.  
The floor for Super-K is about two orders of magnitude below the current limit. Given the low event rate and the \eav\ background, it is unlikely that this floor will be reached.  
For IceCube, the situation is more interesting, as the sensitivity floor is only about one order of magnitude below the current limit at high DM masses.

To cross check, we also estimate the neutrino floor for IceCube with the ``realistic'' approach, using the effective area provided in Ref.~\cite{Aartsen:2016zhm}, which is optimized for solar DM searches and covers roughly $30-3000$\,GeV neutrino energy.  
With this effective area, we find that both the \sav\ background and DM signal reduce by roughly a factor of 10, compared to the ideal case.  
This factor mainly comes from the small signal efficiency factor due to cuts on removing atmospheric muon backgrounds, which could mis-reconstruct and mimic a neutrino event~\cite{Aartsen:2016zhm}.  Other contributions to this factor likely come from various approximations, such as the detector volume and effective area. 
The sensitivity floor obtained with our ideal approach and the realistic case agree well with each other.  This is because the detector efficiency affects both the DM signal and \sav\ background by roughly the same factor.  The small difference at the low mass end of IC is expected from threshold effects.

In Figure~\ref{fig:neutrino_floor}, we also show the strongest direct detection limit currently available, from PICO-60~\cite{Amole:2017dex}. For the $\tau {\bar \tau}$ channel, the solar DM search is more sensitive in most of the mass range.  For the $b {\bar b}$ channel, direct detection experiments are already more sensitive, and are not far from the indirect detection neutrino floor.  (To be clear, direct detection sensitivity is not limited by the indirect detection neutrino floor.)
In all cases, the solar DM search is complementary to direct detection, most notably due to their different dependence on the local DM velocity distribution~\cite{Choi:2013eda}.  

It is informative to compare the neutrino floors between direct detection and indirect detection.  We summarize the results for several representative experimental setups considered in Ref.~\cite{Ruppin:2014bra}.  For heavy targets, such as Ge and Xe, the direct detection floor is higher in general~(even higher for Si targets), which is roughly $10^{-41}\,{\rm cm^{2}}$ at 10\,GeV DM mass and $10^{-40}\,{\rm cm^{2}}$ at 1\,TeV.  This is expected, as heavy targets are more efficient at detecting the background MeV neutrinos through coherent scattering.  In the case of light targets, such as $\rm CF_{3}I$ or $\rm C_{3}F_{8}$, the direct detection neutrino floor is significantly lower, roughly $10^{-43}\,{\rm cm^{2}}$ at both 10\,GeV and 1\,TeV~(even lower if energy information is utilized).  This is lower than the indirect detection neutrino floor for solar DM searches.  Hence, if the indirect detection neutrino floor is reached in the future, a large direct detection experiment with light targets will be needed to reach small SD DM-nucleon cross sections efficiently.  

It is also important to note the subtle differences between the two types of neutrino floors.  The indirect detection neutrino floor can be considered as a ``hard floor''; due to the large theoretical uncertainty of the \sav\ flux, the sensitivity cannot improve once \sav\ events are detected.  Therefore, it important to have a model that can reliably predict the \sav\ flux, taking into account magnetic fields.  If the \sav\ flux is known robustly, then the floor can become ``soft'', meaning that the sensitivity can in principle improve with the square root of the exposure.  It will be difficult to further lower the floor unless good neutrino angular~(good enough to resolve the Sun) and energy resolution are achieved. 

For direct detection, the neutrino floor is already ``soft'', as solar and earth atmospheric neutrino fluxes and their coherent scattering are either well known or can be measured~\cite{Akimov:2017ade}.  The floor can also be lowered with innovative techniques that utilizing energy, timing, and directional information that distinguishes neutrinos from DM, until systematic uncertainties become important. See Ref.~\cite{Battaglieri:2017aum} for a partial collection of ideas.

\subsection{DM Models with Long-Lived Mediators}

In this subsection, we briefly discuss a non-minimal DM scenario that modifies the discussion above. 
If DM annihilates first into a pair of long-lived dark mediators, the neutrinos produced through the delayed decay of those mediators can potentially freely escape the Sun~\cite{Bell:2011sn, Leane:2017vag}.  In this case, it is possible for TeV-scale DM to mimic the high-energy part of the \sav.  
Given that the detection of the TeV \sav\ events are possibly imminent, the neutrino floor for high-mass DM with long-lived mediators may soon be reached. 

To distinguish \sav\ from neutrinos in the long-lived mediator scenario, provided that the mediators decay outside the solar atmosphere, TeV gamma rays~\cite{Batell:2009zp, Leane:2017vag, Arina:2017sng} or electrons~($e^{\pm}$)~\cite{Schuster:2009au, Schuster:2009fc, Ajello:2011dq, Feng:2016ijc} from the Sun may be the key.  
For the long-lived mediator scenario, the gamma-ray and electron flux can be comparable to the neutrino flux, and can be probed by sensitive ground-based~(HAWC~\cite{Abeysekara:2013tka}, LHAASO~\cite{He:2016del}) and space-borne~(Fermi~\cite{Abdollahi:2017kyf}, AMS-02~\cite{PhysRevLett.113.121102}, DAMPE~\cite{ChangJin:550}, CALET~\cite{Torii201155}, etc) detectors.  
However, in the case of cosmic-ray interactions with the Sun, the gamma-ray flux is hugely suppressed by the small angular size of the solar limb~\cite{Zhou:2016ljf}.  
Therefore, multi-messenger GeV-TeV observations of the Sun are important in both understanding the cosmic-ray interactions and general DM searches.


\section{Discussions and Conclusions}
\label{sec:conclusion}

\subsection{ Discussions }
We focus on neutrino-induced muon tracks due to their directionality. 
However, showers induced by electron and tau neutrinos are also powerful signatures for neutrino detection~\cite{Beacom:2004jb}, as they have better neutrino energy resolution and have lower \eav\ backgrounds.  The energy information is important in improving the sensitivity of DM searches~\cite{Rott:2011fh}, and has been demonstrated at low energies in Super-K~\cite{Choi:2015ara}.  
Improved angular resolution for shower events is also expected with KM3NeT~\cite{Adrian-Martinez:2016fdl} in the high-energy regime.
The neutrino flux from DM annihilations has a very different spectrum shape compared to \sav.  
As a result, if showers can achieve a comparable DM sensitivity to muons, they will be important for distinguishing the \sav\ background.   
A shower sensitivity study required detailed understanding of KM3NeT/ARCA, such as the angular and energy resolution at $\sim 100$\,GeV, and is beyond the scope of this work. 

It is interesting to also consider collider probes of the SD DM cross section~(e.g.,~\cite{Goodman:2010ku, Abdallah:2015ter, Zhou:2013fla, Sirunyan:2017hci, Aaboud:2017dor}), given that both direct detection and solar DM searches have neutrino sensitivity floors.  The sensitivity of collider searches is mostly determined by the maximum collision energy and the luminosity of the experiment.  In some cases, collider searches can be more sensitive than both direct and indirect detection, and reach below the neutrino floors~\cite{Harris:2015kda}.   However, comparison with collider sensitivities also involves substantial model-dependent uncertainties~\cite{Buchmueller:2013dya, Buchmueller:2014yoa}.  Thus, direct detection, indirect detection, and collider searches should be considered as complementary probes~\cite{Arrenberg:2013rzp}. 

Due to the small number of events, it is unlikely that Super-K and similarly sized detectors, such as DUNE~\cite{An:2015jdp}, JUNO~\cite{An:2015jdp}, and Jinping~\cite{JinpingNeutrinoExperimentgroup:2016nol} will reach the neutrino floor~(or detect \sav.)  For larger future neutrino detectors, such as Hyper-Kamiokande~\cite{Abe:2011ts, Abe:2016ero}~($\simeq 0.5$\,MT) and PINGU/KM3NeT-ORCA~\cite{Aartsen:2014oha, Adrian-Martinez:2016fdl}~($\simeq 5$\,MT), $\simeq 1$ and $\simeq 10$ events may be present in the detector with 10 years of live time.  The challenge will be to reduce the \eav\ background uncertainty to a similar level. For IceCube-Gen2~\cite{Aartsen:2014njl}, the 10 gigaton extension of IceCube, a muon energy threshold down to at least 10\,TeV is required to for it to be sensitive to \sav. 

\subsection{ Conclusions }

To conclude, in this work we consider the detection of \sav\ neutrinos and their implications for solar DM searches. 

We show that in the multi-TeV regime, where muons can be isolated using their energy loss, \eav\ background can be significantly reduced.  Importantly, \sav\ could be detectable using $\gtrsim 1$\,TeV muons and 10 years of livetime in IceCube and KM3NeT. 
This would help understand cosmic-ray interactions in the solar atmosphere. 
These events cannot be mimicked by standard WIMP scenarios due to neutrino absorption in the Sun.  
However, DM with long-lived mediators could mimic these high-energy neutrinos.    
If these events are detected, TeV gamma rays or electrons could be important diagnostic tools.  

For the sub-TeV regime and considering IceCube,  \sav\ events are indistinguishable with solar DM signals due to the lack of energy resolution.  Therefore, \sav\ constitute a neutrino sensitivity floor, which is only about one order of magnitude below the current IceCube limit.  To breach the floor would require an accurate model of \sav\ that includes magnetic-field effects, or new detector and analysis techniques that can better reconstruct neutrino energy and direction.  Even then, the DM sensitivity can only improve with the square root of the exposure. 

At lower energies and considering Super-K,  the number of expected \sav\ events are much less than one and the neutrino floor is about two orders of magnitude below the current limit.  Therefore, there is still a large discovery potential for low-threshold neutrino telescopes, provided that large detector mass and exposure are available, for example with Hyper-Kamiokande~\cite{Abe:2011ts, Abe:2016ero}.

Solar DM annihilation searches are more sensitive to the SD WIMP-proton cross section than direct searches if the annihilation channel is hard, although the opposite is true for soft annihilation spectra.
If the solar DM sensitivity floor is reached by neutrino telescopes, a large direct detection experiment with light targets may be able to reach one order of magnitude lower in cross section, until it is limited by MeV neutrino backgrounds. 

\vspace{0.2cm}
{\bf Note added:}  During the final stage of this work, Ref.~\cite{Arguelles:2017eao} appeared on arXiv.  In addition, Ref.~\cite{Edsjo:2017kjk} appeared shortly after ours.   Refs.~\cite{Arguelles:2017eao, Edsjo:2017kjk} both provided an updated no-magnetic field \sav\ flux calculation and estimated the neutrino floor.  In comparison, we focus more on the detectability of \sav\ and the implications of the neutrino floor.   Our results agree well with those of Refs.~\cite{Arguelles:2017eao, Edsjo:2017kjk}.

\section*{Acknowledgments}

We thank Carlos Arguelles, David Berge, Francesco Capozzi, Joakim Edsjo, Benjamin Jones, Spencer Klein, Rebecca Leane, Shirley Li, Piotr Mijakowski, Louis Strigari, and Bei Zhou for helpful comments and discussions.
KCYN is supported by a Croucher Fellowship and a Benoziyo Fellowship, and was partially supported by NSF Grant PHY-1404311 awarded to JFB. 
JFB is supported by NSF Grant PHY-1404311. 
KCYN and AHGP were supported by NASA grant NNX13AP49G awarded to AHGP and CR.  
CR  acknowledges  support  from  the  Korea  Neutrino Research Center which is established by the National Research Foundation of Korea (NRF) grant funded by the Korea government (MSIP) (No.  2009-0083526) and Basic Science Research Program NRF-2016R1D1A1B03931688.

\bibliographystyle{h-physrev}
\bibliography{reference}

\end{document}